# Coordinated Transmit and Receive Processing with Adaptive Multi-stream Selection

HongSun An
*Student Member, IEEE*
The Graduate School of IT & T
Inha University
Incheon, Korea
ahs3179@gmail.com

Manar Mohaisen
*Student Member, IEEE*
The Graduate School of IT & T
Inha University
Incheon, Korea
manar.subhi@gmail.com

DongKeol Han

The Graduate School of IT & T
Inha University
Incheon, Korea
xellos1982@naver.com

KyungHi Chang
*Senior Member, IEEE*
The Graduate School of IT & T
Inha University
Incheon, Korea
khchang@inha.ac.kr

*Abstract*—In this paper, we propose an adaptive coordinated Tx-Rx beamforming scheme for inter-user interference cancellation, when a base station (BS) communicates with multiple users that each has multiple receive antennas. The conventional coordinated Tx-Rx beamforming scheme transmits a fixed number of data streams for each user regardless of the instantaneous channel states, that is, all the users, no matter they are with ill-conditioned or well-conditioned channels, have the same number of data streams. However, in the proposed adaptive coordinated Tx-Rx beamforming scheme, we adaptively select the number of streams per user to solve the inefficient problem of the conventional coordinated Tx-Rx beamforming scheme. As a result, the BER performance is improved. Simulation results show that the proposed algorithm outperforms the conventional co-ordinated Tx-Rx beamforming algorithm by 2.5dB at a target BER of $10^{-2}$.

*Keywords-Multi-stream selection; interference cancellation; block diagonalization; coordinated transmit and receive processing; multi-user MIMO.*

## I. Introduction

Multiple-input multiple-output (MIMO) system is an attractive technology due to its capability to linearly increase the system throughput without requiring additional spectral resources [1]. When the channel state information (CSI) is available at the transmitter side by means of feedback, precoding techniques can be applied to improve bit error rate (BER) performance. Dirty paper coding (DPC) is well known as optimal capacity precoding approach for downlink multi-user MIMO (MU-MIMO) communication systems. However, DPC is difficult to apply in practical systems because it is a complex nonlinear scheme. Therefore, precoding scheme with low complexity has received considerable attention.

Block diagonalization (BD) scheme, which decompose the MU-MIMO link into parallel single-user MIMO (SU-MIMO) links, supports multi-stream transmission [2]. BD uses a precoding matrix that ensures zero inter user interference (IUI), where consequently users' data are processed in parallel leading to a reduction in the processing time at the BS side. However, there are drawbacks related to BD scheme. These include the number of users and receive antennas [3].

Coordinated Tx-Rx beamforming scheme was introduced to solve the constraints of BD [3], [4]. In this paper, we propose adaptive multi-stream selection to get more improved performance than the conventional coordinated Tx-Rx beamforming. The proposed scheme allocates the number of data streams for each user based on the consideration of the instantaneous channel state.

The remainder of this paper is organized as follows. System model is given in Section II. In Section III, we describe IUI cancellation schemes. In Section IV, proposed adaptive multi-stream selection is explained. Simulation results and discussions are given in Section V, and conclusions are drawn in Section VI.

## II. System Model

We consider a downlink MU-MIMO system where a base station employing $N_T$ transmit antennas communicates simultaneously with $K$ multiple-antenna non-cooperative mobile stations. We assume that the transmitted data vector for each user is indicated as $\mathbf{D}_k$ ($k = 1, 2, …, K$), and the data symbol vector $\mathbf{D}_k$ is transmitted simultaneously to $k$ users after the pre-processing at the transmitter side. The system model is shown in Fig. 1.

We assume a flat fading channel, where all the elements in the $N_R \times N_T$ channel matrix $\mathbf{H}$ keep constant over one frame duration. The $N_{Rk} \times N_T$ channel $\mathbf{H}_k$ means the channel for user $k$. The transmitter has the full knowledge of the CSI, where the constraint on the total transmit power equals $N_T$.

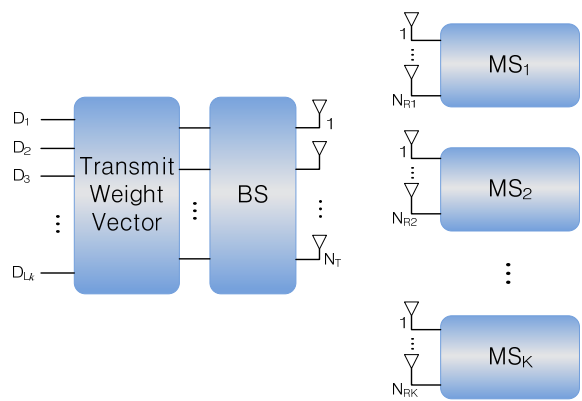

Figure 1. MU-MIMO system with precoding.

This work was supported by the Korea Science and Engineering Foundation (KOSEF) grant funded by the Korea government (MEST) (No. R01-2008-000-20333-0).

Let the $\mathbf{T}_k$ be the $N_T \times L_k$ transmit weight vector, then user $k$ receives $L_k$ data streams through the $N_{Rk}$ receive antennas. Therefore the $N_{Rk} \times 1$ received signal vector is expressed as following:

$$\mathbf{y}_k = \mathbf{H}_k \sum_{i=1}^{K} \mathbf{T}_i \mathbf{D}_i + \mathbf{n}_k, \quad (1)$$

where $\mathbf{n}_k$ is the $N_{Rk} \times 1$ Gaussian noise vector with elements having zero mean and variance of $\sigma_n^2$.

### III. IUI CANCELLATION SCHEMES FOR MU-MIMO SYSTEMS

#### A. BD Algorithm

The IUI can be fully canceled out by using the BD algorithm, where the MU-MIMO channel is transformed into parallel SU-MIMO channels [5]. The purpose of BD algorithm is to find $\mathbf{B}$ such that

$$\mathbf{HB} = \begin{bmatrix} \mathbf{H}_1 \mathbf{B}_1 & \mathbf{H}_1 \mathbf{B}_2 & \cdots & \mathbf{H}_1 \mathbf{B}_K \\ \mathbf{H}_2 \mathbf{B}_1 & \mathbf{H}_2 \mathbf{B}_2 & \cdots & \mathbf{H}_2 \mathbf{B}_K \\ \vdots & \vdots & \ddots & \vdots \\ \mathbf{H}_K \mathbf{B}_1 & \mathbf{H}_K \mathbf{B}_2 & \cdots & \mathbf{H}_K \mathbf{B}_K \end{bmatrix}$$

$$= \begin{bmatrix} \mathbf{H}_{eff,1} & \mathbf{0}_{N_R} & \cdots & \mathbf{0}_{N_R} \\ \mathbf{0}_{N_R} & \mathbf{H}_{eff,2} & \cdots & \mathbf{0}_{N_R} \\ \vdots & \vdots & \ddots & \vdots \\ \mathbf{0}_{N_R} & \mathbf{0}_{N_R} & \cdots & \mathbf{H}_{eff,K} \end{bmatrix}, \quad (2)$$

$$\mathbf{H}_i \mathbf{B}_k = 0 \text{ for all } i \neq k \text{ and } 1 < i, \quad (3)$$

where $\mathbf{0}_{NR}$ is the $N_R \times N_R$ zero matrix and $\mathbf{H}_{eff,k} = \mathbf{H}_k \mathbf{B}_k$ is the effective channel matrix of user $k$ after the BD [6]. To this end, we define the matrix

$$\tilde{\mathbf{H}}_k = \begin{bmatrix} \mathbf{H}_1^T & \cdots & \mathbf{H}_{k-1}^T & \mathbf{H}_{k+1}^T & \cdots & \mathbf{H}_K^T \end{bmatrix}^T, \quad (4)$$

which is obtained by simply removing the channel matrix of user $k$ from the system channel matrix $\mathbf{H}$. The singular value decomposition (SVD) of $\tilde{\mathbf{H}}_k$ is formed as follows:

$$\tilde{\mathbf{H}}_k = \tilde{\mathbf{U}}_k \tilde{\mathbf{\Sigma}}_k \begin{bmatrix} \tilde{\mathbf{V}}_k^1 & \tilde{\mathbf{V}}_k^0 \end{bmatrix}^H, \quad (5)$$

where the columns of $\tilde{\mathbf{V}}_k^0$ are the right singular vectors corresponding to the zero singular values of $\tilde{\mathbf{H}}_k$. Since the columns of $\tilde{\mathbf{V}}_k^0$ lead to zero IUI, they will be potential beamformers for user $k$. Therefore, a linear combination of these vectors is found to form the beamforming matrix $\mathbf{B}_k$. To accomplish this, the SVD of $\mathbf{H}_k \tilde{\mathbf{V}}_k^0$ is formed as follows:

$$\mathbf{H}_k \tilde{\mathbf{V}}_k^0 = \begin{bmatrix} \mathbf{U}_k^1 & \mathbf{U}_k^0 \end{bmatrix} \begin{bmatrix} \mathbf{\Sigma}_k & 0 \\ 0 & 0 \end{bmatrix} \begin{bmatrix} \mathbf{V}_k^1 & \mathbf{V}_k^0 \end{bmatrix}^H, \quad (6)$$

where $\mathbf{H}_k \tilde{\mathbf{V}}_k^0$ is considered to be full-rank. $\mathbf{B}_k$ is then equal to $\tilde{\mathbf{V}}_k^0 \mathbf{V}_k^1$, and finally, the transmit beamforming matrix $\mathbf{B}$ is given by

$$\mathbf{B} = \begin{bmatrix} \tilde{\mathbf{V}}_1^0 \mathbf{V}_1^1 & \cdots & \tilde{\mathbf{V}}_K^0 \mathbf{V}_K^1 \end{bmatrix}. \quad (7)$$

#### B. Coordinated Tx-Rx Processing

There are some limitations for BD algorithm as mentioned before. The coordinated Tx-Rx processing is based on channel decomposition, which was proposed to solve the constraints of BD. The process of coordinated Tx-Rx beamforming is divided into two steps. In the first step, we assume that there is a $N_{Rk} \times L_k$ pre-receiver filter matrices $\mathbf{R}_k$ at the receiver side of each user. The $L_k \times 1$ received signal through pre-receiver is expressed as following:

$$\hat{\mathbf{y}}_k = \mathbf{R}_k^H \mathbf{y}_k. \quad (8)$$

By considering the pre-receiver as part of the channel, we define a new block matrix $\mathbf{H}_S$.

$$\mathbf{H}_S = \begin{bmatrix} \mathbf{R}_1^H \mathbf{H}_1 \\ \vdots \\ \mathbf{R}_K^H \mathbf{H}_K \end{bmatrix} \Rightarrow \mathbf{H}_{S_k} = \begin{bmatrix} \mathbf{R}_1^H \mathbf{H}_1 \\ \vdots \\ \mathbf{R}_{k-1}^H \mathbf{H}_{k-1} \\ \mathbf{R}_{k+1}^H \mathbf{H}_{k+1} \\ \vdots \\ \mathbf{R}_K^H \mathbf{H}_K \end{bmatrix} \quad (9)$$

$$= \begin{bmatrix} \tilde{\mathbf{U}}_k & \mathbf{U}_k \end{bmatrix} \cdot \begin{bmatrix} \mathbf{\Sigma} & 0 \\ 0 & 0 \end{bmatrix} \cdot \begin{bmatrix} \tilde{\mathbf{V}}_k^H \\ \mathbf{V}_k^H \end{bmatrix},$$

By applying BD algorithm for the new block matrix $\mathbf{H}_S$, we calculate the transmit precoding matrix $\mathbf{T}_k$ such that

$$\mathbf{T}_k = \mathbf{V}_k \mathbf{A}_k. \quad (10)$$

In (9) and (10), $\mathbf{V}_k$ is an orthogonal column satisfying following condition:

$$\mathbf{R}_i^H \mathbf{H}_i \mathbf{V}_k = 0 \text{ for } i \neq k. \quad (11)$$

Therefore, the received signal after pre-receiver can be expressed as following:

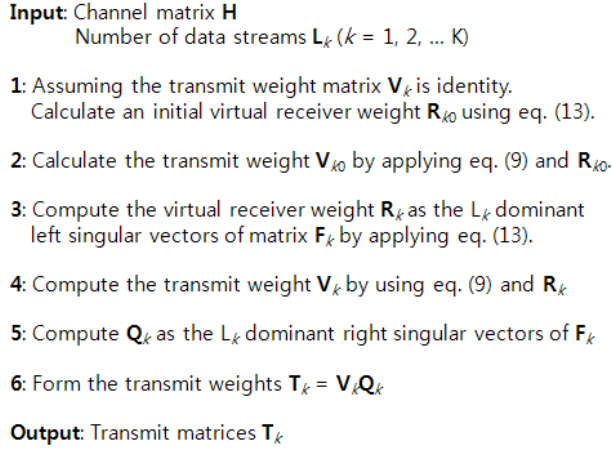

Figure 2.  Coordinated Tx-Rx beamforming scheme.

$$\begin{aligned}\hat{\mathbf{y}}_k &= \mathbf{R}_k^H \mathbf{H}_k \mathbf{T}_k \mathbf{b}_k + \mathbf{R}_k^H \mathbf{n}_k \\ &= \mathbf{R}_k^H \mathbf{H}_k \mathbf{V}_k \mathbf{A}_k \mathbf{b}_k + \mathbf{R}_k^H \mathbf{n}_k.\end{aligned} \quad (12)$$

In the second step, all transmit weight matrices $\mathbf{T}_k$ are calculated at the transmitter side. In (12) the effective channel matrix for user $k$ and $\mathbf{A}_k$ are expressed as

$$\mathbf{F}_k = \mathbf{H}_k \mathbf{V}_k = \begin{bmatrix} \mathbf{R}_k & \mathbf{U}' \end{bmatrix} \cdot \begin{bmatrix} \Sigma_k & \mathbf{0} \\ \mathbf{0} & \Sigma \end{bmatrix} \cdot \begin{bmatrix} \mathbf{Q}_k^H \\ \mathbf{V}_k'^H \end{bmatrix} \quad (13)$$

$$\mathbf{A}_k = \mathbf{Q}_k \mathbf{P}_k^{1/2}, \quad (14)$$

where $\mathbf{Q}_k$ has the orthogonal columns, $\mathbf{P}_k$ is diagonal matrix with positive elements. The diagonal elements of $\mathbf{P}_k$ denote the transmit power for each data-stream of user $k$, and $P_k$=trace($\mathbf{P}_k$) is the total transmit power for user $k$. Therefore, the total transmit power equals to the summation of $P_k$. In this paper, we allocate same amount of transmit power for each data-stream.

The description of coordinated Tx-Rx beamforming scheme is given in Fig. 2.

## IV. PROPOSED ADAPTIVE MULTI-STREAM SELECTION SCHEME

The coordinated Tx-Rx beamforming scheme gives an additional degree of freedom (DoF) over BD algorithm by allocating the different number of transmitted data-streams for each user. Since the additional DoF can offer a gain of signal to noise ratio (SNR), an optimal number of data-streams can be allocated for each user to improve the BER performance.

The conventional coordinated Tx-Rx beamforming scheme assumes that a BS transmits a fixed number of data-streams for each user [3], [4]. In this paper, we propose the adaptive multi-stream selection scheme by considering the instantaneous channel state. The proposed scheme assumes that the receiver use linear detection schemes to detect the transmitted signals.

When zero-forcing (ZF) scheme, which detects the transmitted signal by multiplying the weight vector, is employed, the procedure can be expressed as following:

$$\begin{aligned}\mathbf{y} &= \mathbf{H}\mathbf{x} + \mathbf{n} \\ \mathbf{W}_{ZF} &= (\mathbf{H}^H \mathbf{H})^{-1} \mathbf{H}^H \\ \mathbf{W}_{ZF} &= (\mathbf{H}^H \mathbf{H})^{-1} \mathbf{H}^H \mathbf{H}\mathbf{x} + (\mathbf{H}^H \mathbf{H})^{-1} \mathbf{H}^H \mathbf{n} \\ &= \mathbf{x} + (\mathbf{H}^H \mathbf{H})^{-1} \mathbf{H}^H \mathbf{n},\end{aligned} \quad (15)$$

where, $\mathbf{H}^H$ is Hermitian transpose matrix. In ZF scheme, the performance degradation is generated by noise amplification as given in the second term of (15) when the channel is ill-conditioned.

In sight of QR-decomposition (QRD) of the channel matrix, the noise amplification factor is given as a function of the matrix $\mathbf{R}$ as follows:

$$\begin{aligned}\mathrm{Tr}\{(\mathbf{H}^H \mathbf{H})^{-1}\} &= \mathrm{Tr}\{(\mathbf{Q}\mathbf{R}\mathbf{R}^H \mathbf{Q}^H)^{-1}\} \\ &= \mathrm{Tr}\{(\mathbf{R}^{-H} \mathbf{R}^{-1})\}.\end{aligned} \quad (16)$$

Let $\mathbf{A}=\mathbf{R}^{-1}$, then, due to the triangular structure of $\mathbf{R}$

$$\begin{aligned}\mathrm{Tr}\{(\mathbf{R}^{-H} \mathbf{R}^{-1})\} &= \sum_{i=1}^M \sum_{j=1}^M \|A_{i,j}\|^2 \\ &= \sum_{i=1}^M \frac{1}{R_{i,i}^2} + \sum_{i=1}^M \sum_{j=i+1}^M \|A_{i,j}\|^2 \\ &\geq \sum_{i=1}^M \frac{1}{R_{i,i}^2},\end{aligned} \quad (17)$$

where, $R_{i,i}$ represents the entries on the diagonal of $\mathbf{R}$. The equality in (17) is satisfied iff. $\mathbf{H}$ is an orthogonal matrix, i.e., $\mathbf{R}$ is diagonal.

Therefore, a sub-optimum subset of the number of multi-stream for each user can be obtained by minimizing $\sum_{i=1}^M 1/R_{i,i}^2$. This can be done by maximizing the diagonal elements of $\mathbf{R}$. For that, QRD-based data-stream selection schemes are not optimum due to discarding the second summation of (17).

### A. Proposed adaptive multi-stream selection scheme

At the first stage of the proposed scheme, sorted-QRD (SQRD) algorithm [8] is applied to the Hermitian transpose matrix $\mathbf{H}^H$ of channel matrix $\mathbf{H}$. Then, the column with the largest power, i.e. the column leading to the lowest noise amplification, is considered leading to reduction in $1/R_{i,i}^2$. The remaining (N-1) columns of the matrix $\mathbf{Q}$ are then orthogo-nalized and the one leading to the lowest noise amplification is selected for the second iteration. The proposed algorithm is processed until obtaining the first subset of the overall number of the transmitted data-stream. Also, the metric $D = \sum_{i=1}^{M} 1/R_{i,i}^2$ is calculated successively, and assigned to temporary threshold value. At the second stage of the proposed scheme, the column with second larger power is selected at the first iteration of the SQRD, where the remaining columns are orthogonalized and the corresponding accumulative metric is calculated. At any iteration, if $D$ exceeds the already-found best threshold value, the stage is finished and the procedure moves to the next stage. If the determined best multi-stream subset through this procedure, which leads to the minimum noise amplification, does not allocate at least one data-stream for all users, the proposed algorithm selects the next optimal subset which support at least one data-stream for each user. Therefore, the proposed algorithm considers the fairness among the users as well as improves the BER performance. Finally, if we determine the number of multi-stream for each user, conventional coordinated Tx-Rx beamforming is applied with the determined number of multi-stream.

Fig. 3 gives the detailed description of the proposed adaptive data-stream selection scheme.

## V. SIMULATION RESULTS

In this Section, we evaluate the bit error rate performance of the proposed adaptive coordinated Tx-Rx beamforming scheme. We consider both 8×8 SU-MIMO system and 8×(2, 2, 2, 2), 8×(4, 4, 4, 4) MU-MIMO system, where the BS has 8 transmit antennas and simultaneously communicates with 2 and 4 multiple-antenna non-cooperating users. The transmitted symbols are drawn from a 4QAM constellation set. The BS has perfect knowledge of CSI by means of feedback from the users. We assume that the conventional coordinated Tx-Rx beamforming scheme uses 2 data-streams for each user, and adaptive coordinated Tx-Rx beamforming scheme uses the determined number of data-streams by considering channel states.

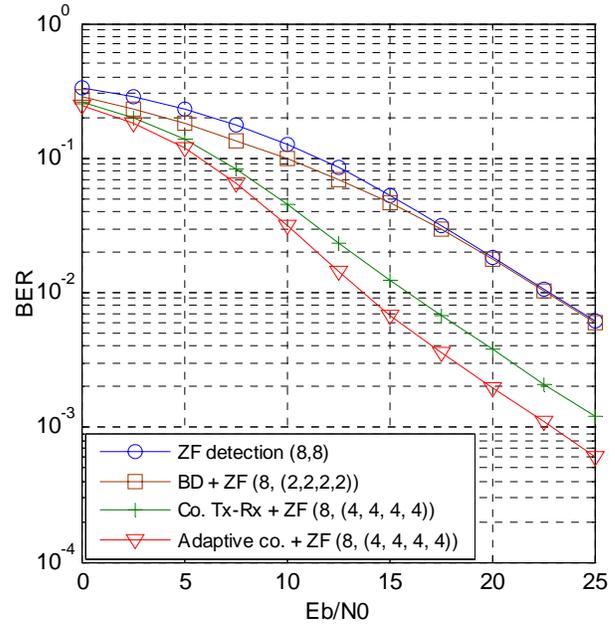

Figure 4. BER performance comparison among IUI cancellation schemes using ZF detection.

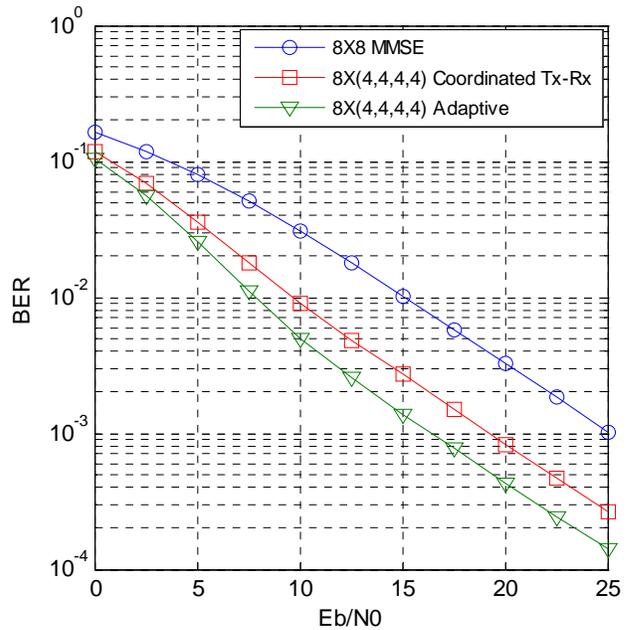

Figure 5. BER performance comparison among IUI cancellation schemes using MMSE detection.

Figure 3. Proposed adaptive multi-stream selection scheme.

Fig. 4 depicts the comparison of BER performance among different IUI cancellation schemes when all the receivers adopt ZF detection scheme, and Fig. 5 shows the same scenario when all the receivers adopt MMSE detection scheme. From the simulation results, we can find that the proposed scheme outperforms the conventional coordinated Tx-Rx beamforming by about 2.5dB at a target BER of $10^{-2}$, when the total number of the transmitted data-stream is fixed to 8.

## VI. CONCLUSIONS

In this paper, we proposed an adaptive coordinated Tx-Rx beamforming scheme for IUI cancellation in MU-MIMO systems with multiple-antenna users. The proposed algorithm allocates the different number of data-streams for each user by considering the instantaneous CSI. And the proposed algorithm also considers fairness, which means each user has at least a single data-stream. Consequently, the adaptive coordinated Tx-Rx processing considerably improves the BER performance. On the assumption that each receiver adopts the linear detection schemes, simulation results show that the proposed scheme outperforms conventional coordinated Tx-Rx beamforming about 2.5dB at the target BER of $10^{-2}$.